\documentclass[conference]{IEEEtran}
\newtheorem{definition}{Definition}

\usepackage{amsmath}
\usepackage{amssymb}
\usepackage{tikz}
\usetikzlibrary{bayesnet}

\usepackage{array}

\usepackage{graphicx}
\usepackage[caption=false,font=footnotesize]{subfig}

\usepackage{dblfloatfix}

\usepackage{url}

\begin{document}

\title{Realistic Traffic Generation for Web Robots}
\author{\IEEEauthorblockN{Kyle~Brown}
\IEEEauthorblockA{ Department of Computer Science and Engineering \\ Kno.e.sis Research Center\\
Wright State University, Dayton, OH, USA\\
Email: brown.718@wright.edu}
\and
\IEEEauthorblockN{Derek~Doran}
\IEEEauthorblockA{ Department of Computer Science and Engineering \\ Kno.e.sis Research Center\\
Wright State University, Dayton, OH, USA\\
Email: derek.doran@wright.edu}
}

\maketitle

\begin{abstract}
Critical to evaluating the capacity, scalability, and availability of web
systems are realistic web traffic generators. Web traffic generation is a
classic research problem, no generator accounts for the characteristics of 
web robots or crawlers that are now the dominant source of traffic to a web server. 
Administrators are thus unable to test, stress, and evaluate how their systems perform
in the face of ever increasing levels of web robot traffic. To resolve this problem, 
this paper introduces a novel approach to generate synthetic 
web robot traffic with high fidelity. It generates traffic that accounts for 
both the temporal and behavioral qualities of robot traffic by statistical and 
Bayesian models that
are fitted to the properties of robot traffic seen in web logs from North America and
Europe. We evaluate our traffic generator by comparing the characteristics of generated
traffic to those of the original data. We look at session arrival rates, inter-arrival
times and session lengths, comparing and contrasting them between generated and real
traffic. Finally, we show that our generated traffic affects cache performance similarly
to actual traffic, using the common LRU and LFU eviction policies.
\end{abstract}

\section{Introduction}
\label{sec:introduction}
Web robots are computer agents that send HTTP requests to Web servers and are
a growing source of traffic on the World Wide Web. A common source of Web
robot traffic are crawlers for search engines, such as Google's 
GoogleBot~\cite{brin1998anatomy}, Microsoft's 
bingbot \footnote{\url{https://www.bing.com/}},
Baidu's BaiduSpider \footnote{\url{http://baidu.com/}}
and Apache Nutch~\cite{khare2004nutch},
an open source search engine framework. There are also more focused crawlers 
that search the web for information about a specific topic, find prices on 
e-commerce sites, collects e-mail address, and discovers broken hyperlinks
\cite{dikaiakos2005investigation,tan2004discovery}. Robots can also be malicious,
attempting to exploit vulnerabilities on a Web server or taking part in a
distributed denial of service (DDoS) attack with other robots in a botnet~
\cite{holz2005short}.

In the past few years, multiple studies have found that Web robots make up the majority
of HTTP traffic. A 2015 industry report suggests that robots constitute 49.5\% 
of all HTTP requests to popular Web sites on the Internet \cite{zeifman2015bot}, while 
a recent academic report suggests a lower bound for this number to be between 50 and 
60\%~\cite{rude2015request, doran2013comparison, sun2010ethicality}.
The rise in robot traffic to these levels may be attributed to many factors, including 
modern software frameworks that make it easy to develop and test web robots (e.g., Python BeautifulSoup
or Apache Nutch) or commonplace social media linking, commenting, and sharing functions that encourage
real-time engagement (hence requiring aggresive crawling to stay up to date with the latest
information shared). 

The high levels of Web robot
traffic faced by web servers can significantly hinder their performance. Server-side web caches,
for example, are especially vulnerable to web robots~\cite{almeida2001analyzing}
as their requests can evict resources likely to be requested by humans, or admit resources
seldom asked for by humans. While the Robots
Exclusion Protocol (REP) allows 
web server administrators some control over robot behavior~\cite{sun2007large,kolay2008larger},
it is an unenforced advisory that robots can violate without concequence~\cite{giles2010measuring}.
Web servers must therefore be expected to handle traffic from \emph{unethical}
robots which do not adhere to the REP. 

Existing approaches to Web traffic generation tend to be based on models of human behavior -
predicting when a person will click on a link and which links they are most likely to 
click next~\cite{liu2001traffic}. Others attempt to model the global properties of Web traffic
which are shared between humans and robots~\cite{choi1999behavioral,mah1997empirical}. These
studies were also done in the late 1990s and early 2000s, and in the years after the structure
and size of the Internet have changed greatly~\cite{ihm2011towards}. Therefore, in order to
understand how modern Web robot traffic impacts server performance, a new approach to traffic
simulation is necessary.

This paper introduces a system for generating Web robot requests that resemble
the pattern of requests observed on actual Web servers. Our system combines
models of the temporal patterns of Web robot behavior with existing models
which predict the subdirectories and resources robots make requests in. We
compare the characteristics of traffic generated by our system with actual
traffic we are trying to model, including the average session length, time
between sessions, and time between requests within a session. We compare the
effect of generated traffic and the effect of actual traffic on two common
caching algorithms, least recently used (LRU) and least frequently used (LFU),
showing that the generated traffic has a similar impact on cache performance.

The remainder of the paper is as follows: Section~\ref{sec:related_work} provides
references to previous research on Web robot behavior, cache evaluation, and
simulation of HTTP traffic. Section~\ref{sec:methodology} introduces the generative
models used to simulate Web robot traffic. Section~\ref{sec:evaluation} compares our
generated traffic to actual Web robot traffic and how each affects the 
performance of common caching algorithms.

\section{Related work}
\label{sec:related_work}
The number of methods for generating web traffic is extensive, but most were
developed over a decade ago~\cite{liu2001traffic,mah1997empirical,choi1999behavioral}. 
Different traffic generators emphasize
different qualities of the users and their session behaviors. For example, 
Liu \textit{et al.} presented a model of Web traffic with applications to the
performance evaluation of Web servers~\cite{liu2001traffic} involving navigational
patterns through hypertext They include a
traffic model at the session level which describes the arrival and browsing
behaviors of users, and describe WAGON, a graphical software interface for
generating traffic and evaluating how it impacts an Apache Web Server.
Mah developed an empirical model based on HTTP packet traces which can be used
for simulations~\cite{mah1997empirical}. The work uses packet traces in order to model
the behavior of a single user across multiple websites, which is much more
difficult to do if server logs are used. This model is more suitable for testing
higher-scale properties of the Web, such as Internet routing and bandwidth
usage, rather than the impact of the traffic on a single server-side cache.
Choi and Limb developed a behavioral model of Web traffic which includes the
effect of caches~\cite{choi1999behavioral}. However, this approach is based off
of the \texttt{last-modified} and \texttt{if-modified-since} fields in the HTTP
header, and not necessarily any server-side caches.

The traffic generator we propose is made with an eye towards supporting performance
evaluation on web servers as it faces varying levels of web robot traffic.
This class of traffic has seen only a small amount of study
over the last two decades. Almeida \textit{et al.} studied the characteristics of Web robot traffic and
their impact on server-side caching~\cite{almeida2001analyzing}. Their results
indicate that Web robot traffic violates the locality assumptions that hold
for human traffic. They conclude that in order to improve performance in the
face of increased Web robot traffic, servers should adopt different caching
policies for human and Web robot requests. Marshall and Roadknight analyzed the
impact a single human user can have on a cache~\cite{marshall1998linking}. Their
study shows that users can have varying hit rates, meaning some may degrade
performance more than others.
Doyle \textit{et al.} describe what they call the \emph{trickle-down} effect the
most popular resources on a Web server have on caching~\cite{doyle2002trickle}.
This process involves upstream caches storing the most popular resources, while
off-loading less popular resources and dynamically generated resources to
downstream sources. These less popular and dynamic resources can negatively impact
the performance of down-stream caches, even while up-stream caches maintain good
performance.

Our work differs from traffic generators in two ways: it is meant specifically for
generating Web robot traffic, and it makes less assumptions about the browsing behavior
of agents. Previous traffic generation models such as those in \cite{liu2001traffic} tend
to assume that agents follow the hyperlink structure of a website; with human agents, this
is often the case, but it may not be so for Web robots. Our traffic generation approach
also generates Web robot traffic that has a similar effect on Web server (proxy)
caches as actual Web robot traffic, whereas other models seek to imitate the overall impact
performance has on the Web server itself.

\section{Methodology}
\label{sec:methodology}
\begin{figure}[t]
	\vspace{-10px}
	\centering
	\includegraphics[width=0.5\textwidth]{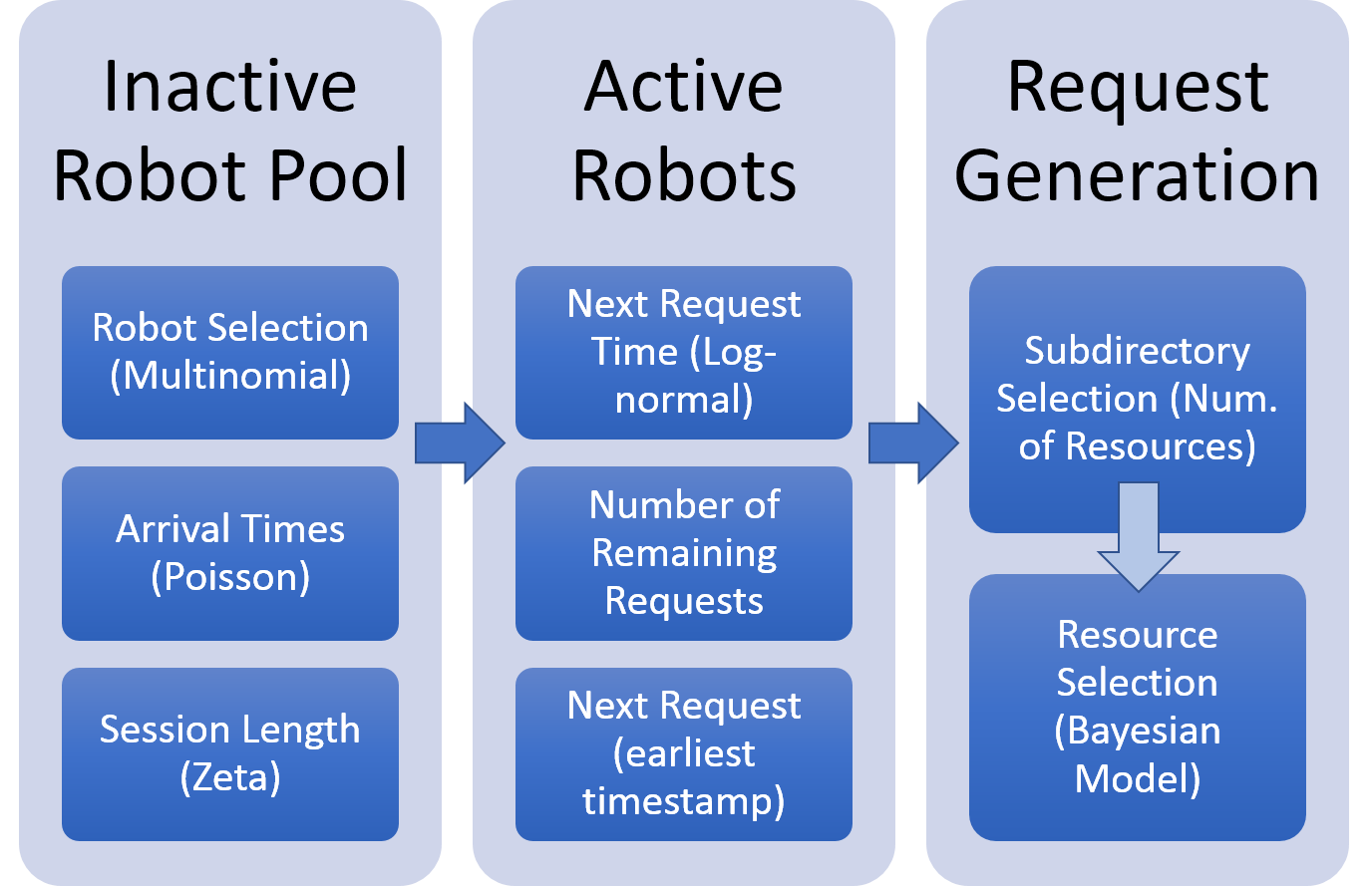}
	\vspace{-10px}
	\caption{High-level description of the traffic generator}
	\label{fig:tgdiagram}
	\vspace{-15px}
\end{figure}

This section presents the methodology behind our Web robot traffic generator. The three main
components of the model can be seen in Figure~\ref{fig:tgdiagram}: 
~(i) an inactive robot pool from which new sessions are drawn;~(ii) an active robot
pool, from which the robot making the next request is chosen; and~(iii) a request generator, which
chooses the subdirectory and resource which is to be requested next. Whenever a new session
is to be created, a robot is chosen from the inactive robot pool by sampling from a
categorical distribution. Each robot in the active robot pool is associated with a session,
which keeps track of the time the robot became active and the session length, the number
of requests the robot will make before becoming inactive. The resource generation components
are responsible for choosing which subdirectory the next request will be made in, and then
which resource within the subdirectory is to be requested, as well as the inter-arrival time,
which is the time between the last request made by the robot making the next request.

In the process of generating a request, several probability distributions are sampled from.
The parameters of these distributions in practice are fit based on actual Web server logs. This
allows for generating traffic matching the characteristics of a particular server, which can
then be used to test the performance of that server. Furthermore, the parameters can be
tweaked to simulate hypothetical situations such as a huge volume of Web robot traffic, or
robots requesting large resources which can negatively impact server performance.

The generator that we develop models web robot traffic as coming from a 
{\em robot pool}, which is simply a list of sessions
which are still ongoing. The number of sessions at any given time is a 
hyperparameter of our traffic generation simulation, denoted $N$. In practice $N$
is estimated from existing Web server logs as the average over the number of
active sessions as each request is processed. Since the size of the robot pool is
fixed, a new session is added whenever another ends to ensure that there are always
$N$ active sessions. 

A central concept of the model is thus a 
\emph{session}, which represents a sequence of requests made by a single agent
to the same Web server over a period of time. This notion is based off of the
approach used in~\cite{calzarossa2011analysis} and~\cite{sisodia2015comparative}
for finding sessions in Web server logs. We model a session as a tuple
\begin{equation}
S = \left(A, t_0, k\right)
\end{equation}
where $A$ is an identifier of the agent making the requests, $t_0$ is the time
of the first request in the session, and $k$ is the total number of requests to
be made in the session, called the \emph{session length}. Each of these components
are viewed as a random variable which follow distributions as described in the 
sequel. 

\subsection{Session Generation}
\label{sec:session_generation}
Session generation requires a model for the number of new sessions that arrive
over a time period and their length. We use the following definition of a session.

\begin{definition}
Let $S = (r_1, r_2, \dots, r_n)$ be a sequence of $n$ requests for resources $r_i$
made by the same agent, and let $t(r_i)$ be the time that resources $r_i$ was
requested. Given some $T > 0$, we say that $S$ is a \emph{session} with timeout
$T$ if for all $i=2, \dots, n$, we have that $t(r_i) - t(r_{i-1}) \le T$.
\end{definition}

Many empirical studies of Web server traffic suggest that a Poisson process is a 
realistic model of session arrival rates and times~\cite{kingman2005poisson}. 
This means that the
probability of a given number of sessions $j$ appearing in a fixed interval of
time follows a Poisson distribution:
\begin{equation}
j \sim \mathrm{Poisson}(\lambda)
\end{equation}
To sample the time at which sessions begin, we consider the random
variable $\Delta t$, which is the time between two sessions from any robots
that occur in sequence, also known as the session inter-arrival time.
Since $j$ is Poisson distributed, $\Delta t$ follows an Exponential
distribution with the same rate parameter $\lambda$:
\begin{equation}
\Delta t \sim \mathrm{Exponential}(\lambda)
\label{eq:tbs_dist}
\end{equation}
Then the initial time for a new session $t_k$ is generated as follows: first,
sample $\Delta t$ as in~\eqref{eq:tbs_dist}. If $t_{k-1}$ is the start time
of the last session, then
\begin{equation}
t_k = t_{k-1} + \Delta t
\end{equation}
The Poisson distribution used to generate session times is fit using maximum
likelihood estimation (MLE); given a time period $T$ and $n_s$, the number of
sessions observed in that period, the MLE estimate for $\lambda$ is
\begin{equation}
\hat{\lambda} = \frac{n_s}{T}
\end{equation}
To model the distribution of web robot session lengths, we use the Zeta distribution,
which is often called a power-law or Zipf distribution~\cite{clauset2009power}. The
family of power-law distributions are distinguished by having a density or mass function
proportional to some (negative) power of the random variable $x$:
\begin{equation}
f(x) \propto x^{-s}
\end{equation}
The Zeta distribution is a discrete probability distribution with mass function
\begin{equation}
f(k) = \frac{k^{-s}}{\zeta(s)}
\end{equation}
with parameter $s > 1$. Here $\zeta(s)$ is the Riemann zeta function, which
is determined by the infinite series
\begin{equation}
\zeta(s) = \sum_{i=1}^{\infty}i^{-s}
\end{equation}
which converges for real $s > 1$. The session length $k$ is sampled from a
Zeta distribution which describes the session length of all robot sessions:
\begin{equation}
k \sim \mathrm{Zeta}(s)
\end{equation}
There is no closed form solution for the MLE of parameter $s$. We chose to
approximate the solution numerically using methods described 
in~\cite{clauset2009power}. The methods directly maximize the log-likelihood
function using numerical maximization algorithms. If the observed session
lengths are $x_1, \dots, x_n$, then the MLE estimate of $s$ is
\begin{equation}
\hat{s} = \underset{s\in (1,\infty)}{\arg\max}
\left\{-n\ln{\zeta(s)}-s\sum_{i=1}^n \ln{x_i}\right\}
\end{equation}
assuming a minimum value of $x_{min} = 1$.

Once temporal information for a new session has been generated, it must be 
assigned to an inactive robot. This robot is selected from the pool of
inactive robots with a probability determined by how frequently it made
requests in the data used to fit the traffic generator. If there are $R$ 
robots and the $i^{th}$ robot made $n_i$ requests, then we define a vector 
$\vec{\rho}$ of probabilities given by
\begin{equation}
\rho_i = \frac{n_i}{\sum_{j=1}^Rn_j}
\end{equation}
When a robot becomes active, it is removed from $\vec{\rho}$ and the
remaining values are renormalized to sum to one. When it becomes inactive,
it is added back in. Given the vector $\vec{\rho}$ containing only
inactive robots, a new active robot $a$ can be sampled from a Multinomial
distribution:
\begin{equation}
a \sim \mathrm{Multinomial}(1;\vec{\rho})
\end{equation}
Within the generator, there are always $N$ active sessions maintained.
When a robot ends its session, a new one is drawn to fill its place.

\subsection{Request Time Generation}
Next, the generator assigns times to each request in a generated web robot session. 
Based on past studies~\cite{almeida2001analyzing} and empirical measurements
shown in Figures~\ref{fig:pavia_compare_iat} and~\ref{fig:wsu_compare_iat},
a lognormal distribution is used to
generate the time between requests within a session. 
Given $n$ data points
$x_1, \dots, x_n$, maximum likelihood-estimates for the lognormal parameters
$\mu$ and $\sigma$ are
\begin{equation}
\hat{\mu} = \frac{1}{n}\sum_{i=1}^n\log x_i
\end{equation}
and
\begin{equation}
\hat{\sigma}^2 = \frac{1}{n-1}\sum_{i=1}^n\left(\log x_i - \mu\right)^2
\end{equation}
i.e. the mean and standard deviation of the logarithm of the samples.

Lognormally distributed random variables can be generated by 
sampling from a normal and exponentiating the result.
To maintain
a small number of parameters, a single global lognormal distribution is
fit based on the times between requests within sessions for all robots.
Given the initial time of a robot session $t_1$, and the length of the
session $k$, the times of each request are generated as follows.
For $i = 2, \dots, k$, sample a time delta from the global lognormal distribution
\begin{equation}
\tau_i \sim \log \mathcal{N}\left(\mu, \sigma^2\right)
\end{equation}
Then the time of the $m$th request in the session is
\begin{equation}
t_m = t_1 + \sum_{i=2}^m \tau_i
\end{equation}

\subsection{Subdirectory and Resource Generation}
The last components of the traffic generator decides on the
subdirectory of a request and the resource to be requested within it. 
To minimize the number of parameters for this component, we
consider the generation of subdirectories globally (i.e. for all robots), and generation of
resource requests by a robot given the previously generated subdirectory.

The simplest method for generating a subdirectory is to just pick one from all
possible subdirectories at random, with equal probabilities. This is equivalent
to placing a discrete uniform probability distribution over the subdirectories. 
Although this method is simple and easy to implement, it ignores the fact that certain
subdirectories, such as the root directory \texttt{/}, are more likely to be requested from
than others. A factor that can influence which subdirectories are more likely to be requested
from others is the number of resources in the subdirectory. If there are $M$ subdirectories
$d_1, \dots, d_M$, with $d_i$ having $R_i$ resources, then we define a vector 
$\vec{\sigma}$ with components
\begin{equation}
\sigma_i = \frac{R_i}{\sum_{k=1}^M R_k}
\end{equation}
Then whenever we need to generate a new subdirectory $d$ we can sample it from a categorical
distribution:
\begin{equation}
d \sim \mathrm{Multinomial}(1;\vec{\sigma})
\end{equation}
This model privileges subdirectories with a large number of resources; a subdirectory with
the minimum of one resource will be the least likely to be generated, since subdirectories with
no resources are not even considered.

To generate a resource given the robot making the request and subdirectory the request
will be in, we define a probability distribution over the resources in that subdirectory.
This model takes into account the types of resources, which are determined by the file
extension. To fit the parameters of the distribution, we do Bayesian inference using 
maximum \emph{a posteriori} (MAP) estimation. Our priors are chosen to incorporate
information about resource and type frequencies in requests from all robots in the
subdirectory.

We select a resource by turning to a generative Bayesian model. 
Given a vector $\vec{\theta}$
where $\theta_i$ is the probability of requesting a resource with type $i$, draw
a resource type $t$ as
\begin{equation}
t \sim \mathrm{Multinomial}(1;\vec{\theta})
\end{equation}
Given a resource type $t$, there is another vector $\vec{P}_t$, where the $i$-th
component $\vec{P}_{t,i}$ is the probability of requesting the $i$-th resource of type $t$
in the subdirectory, we can draw the resource $r$ as
\begin{equation}
r \sim \mathrm{Multinomial}(1;\vec{P_t})
\end{equation}
\begin{figure}[t]
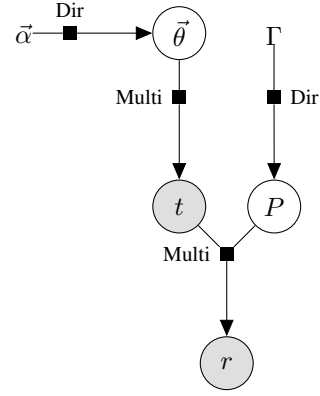

	\centering
	\vspace{-30px}
	\tikz{
		\node[const](alpha){$\vec{\alpha}$};
		\factor[right=of alpha]{a-factor}{Dir}{}{};
		
		\node[latent, right=of a-factor](theta){$\vec{\theta}$};
		\factor[below=of theta]{th-factor}{left:Multi}{}{};
		
		\node[obs, below=of th-factor](t){$t$};
		\factor[below right=of t]{tp-factor}{left:Multi}{}{};
		\node[obs, below=of tp-factor](r){$r$};
		\node[latent, above right=0.4 of tp-factor](p){$P$};
		\factor[right=1.07 of th-factor]{gam-factor}{right:Dir}{}{};
		\node[const, above=0.6 of gam-factor](gamma){$\Gamma$};
		
		\factoredge{alpha}{a-factor}{theta};
		\factoredge{theta}{th-factor}{t};
		\factoredge{t,p}{tp-factor}{r};
		\factoredge{gamma}{gam-factor}{p};
	}
	\vspace{-5px}
	\caption{Bayesian network for the request path model}
	\label{fig:sdm_bayes_net}
	\vspace{-10px}
\end{figure}
When using MAP to estimate the parameters $\vec{\theta}$ and all of the $\vec{P}_t$, we place
Dirichlet conjugate priors~\cite{agresti2005bayesian} on them:
\begin{equation}
\vec{\theta} \sim \mathrm{Dirichlet}(\vec{\alpha})
\end{equation}
and
\begin{equation}
\vec{P}_t \sim \mathrm{Dirichlet}(\vec{\gamma}_t)
\end{equation}

This entire generative process is summarized in Figure~\ref{fig:sdm_bayes_net} using the notation
described in~\cite{dietz2010directed}. Here $P$ stands
for all of the $\vec{P}_t$, and $\Gamma$ stands for all of the $\vec{\gamma}_t$. Given an observed
sequence of $M$ requests $R = (r_1, \dots, r_M)$ with corresponding types $T = (t_1, \dots, t_M)$, 
the data likelihood can be written
\begin{eqnarray}
\mathrm{Pr}(R, T | \vec{\theta}, P) =& \nonumber\\
\exp\left\{
\sum_{j=1}^K\left(m_j\log(\theta_j) + \sum_{l=1}^{R_j}n_{j,l}\log(p_{j,l})\right)\right\}
\end{eqnarray}
where $K$ is the number of resource types, 
$m_j = \sum_{k=1}^M \mathbb{I}\left[t_k = j\right]$ is the number of requests for
resources of type $j$ by the robot, $R_j$ is the number of resources of type  $j$ in the subdirectory,
and $n_{j,l}$ is the number of requests for the $l$-th resource of type $j$. Since we are using
MAP for parameter estimation, we seek values for $\vec{\theta}$ and $P$ that maximize the 
posterior
\begin{equation}
\Pr(\vec{\theta}, P | R, T) = 
\frac{\Pr(R, T | \vec{\theta}, P)\Pr(\vec{\theta})\Pr(P)}{\Pr(R,T)}
\end{equation}
Since we chose conjugate priors, the MAP parameter estimates can be found analytically as
\begin{equation}
\label{eq:res_type_map}
\tilde{\theta}_j = \frac{\alpha_j + m_j - 1}{\alpha + M - 1}
\end{equation} for $1 \le j \le K$ and
\begin{equation}
\label{eq:res_path_map}
\tilde{p}_{j,l} = \frac{\gamma_{j,l} + n_{j,l} - 1}{\Gamma_j + m_j - 1}
\end{equation} for $1 \le j \le K$ and $1 \le l \le R_j$, where
$\alpha = \sum_{j=1}^K \alpha_j$
and $\Gamma_j = \sum_{l=1}^{R_j} \gamma_{j,l}$.

The hyperparameters $\vec{\alpha}$ and $\Gamma$ are chosen based on global statistics;
let $(m_G)_j$ be the number of requests for a resource of type $j$ in the subdirectory across
all robots. Also, let $(n_G)_{j,l}$ be the number of requests for the $l$-th resource of type
$j$ in the subdirectory across all robots. Then for $\alpha > 0$ and $\gamma > 0$, we set
\begin{equation}
\alpha_j = \alpha\frac{(m_G)_j}{\sum_{k=1}^M(m_G)_k}
\end{equation}
and
\begin{equation}
\gamma_{j,l} = \gamma\frac{(n_G)_{j,l}}{\sum_k=1^{R_j}(n_G)_{j,k}}
\end{equation}
The values $\alpha$ and $\gamma$ allow for control over the strength of the prior information; the
larger their values, the more the hyperparameters contribute to the probabilities in 
Equations~\eqref{eq:res_type_map} and~\eqref{eq:res_path_map}. By default, we either set $\alpha$
and $\gamma$ to a constant, or to be related to the number of data points included in the prior
information.

\section{Evaluation}
\label{sec:evaluation}
To evaluate our web robot traffic generator, we fit its parameters to
those of actual traffic and compare the characteristics of the generated traffic
to those of the original traffic. Two datasets were used for evaluation: the first consists
of web robot traffic across every web server at Wright State University\footnote{\url{wright.edu}} 
(WSU) for the months of April, May, and June in 2016. 
The second consists of web robot traffic from the servers of the University of Pavia in 
Italy\footnote{\url{www.unipv.eu/site/en/home.html}} (Univ. of Pavia) 
from December 2013 to May 2014. As the
original datasets were server logs of \emph{all} traffic, we extracted web robot traffic
by comparing the User-Agent HTTP header field with the crowd-sourced database
BotsVsBrowsers~\cite{bvb_site}. 
Each log entry includes at least the time and date the request was observed, 
HTTP request including method, path, and HTTP version, HTTP response
code from the server, and IP address of the requester. 
Other fields which may or may not be present are the response
size, referrer field in the HTTP header, and User-Agent string. Agents are identified
by their User-Agent string, IP address, or both.

Table~\ref{tbl:dataset_summary} summarizes the Web robot
traffic extracted from the datasets. We find that a significant amount of traffic is observed on both servers, exceeding 250,000 robot requests over the past three months, but there are important
differences in their summary statistics. For example, sessions on 
Univ. of Pavia is smaller, with an average length of 9.42 requests, whereas sessions last
34.49 requests on average on WSU. The diversity of robot traffic on WSU is also much stronger,
with nearly twice as many unique robots visiting this server. This may be due to the fact
that the WSU website is much richer in information, having over twice the number of unique
resources available for download. These differences are a positive aspect, 
as it allows us to evaluate if the generator can mimic patterns from two diverse and distinct
sets of robot traffic. 

\begin{table}
\vspace{-10px}
\centering
\caption{Summary of Web Robot traffic datasets}
\begin{tabular}{|c||c|c|}
\hline
\bfseries{Metric} & \bfseries{Univ. of Pavia} & \bfseries{WSU} \\
\hline
Num. Requests & 269,516 & 1,790,036 \\
Num. Sessions & 28,583 & 51,898 \\
Num. Agents & 226 & 576 \\
Num. IP addresses & 3,796 & 9,578 \\
Num. Resources & 59,286 & 145,369 \\
\hline
\end{tabular}
\label{tbl:dataset_summary}
\vspace{-10px}
\end{table}
To evaluate the quality of the generated traffic, we compare the 
 distribution of the time between new session arrivals across all Web robots, 
 the inter-arrival times between requests within the same session, and the
 distribution of session lengths across all robots. 
Finally, we implement a simple cache with least-frequently used (LFU) and least-recently
used (LRU) eviction policies and compare the impact of generated and original traffic
on these caching policies.

\subsection{Time Between Sessions}
\begin{figure}[t]

	\centering
	\subfloat[Univ. of Pavia] {
		\includegraphics[width=0.25\textwidth]{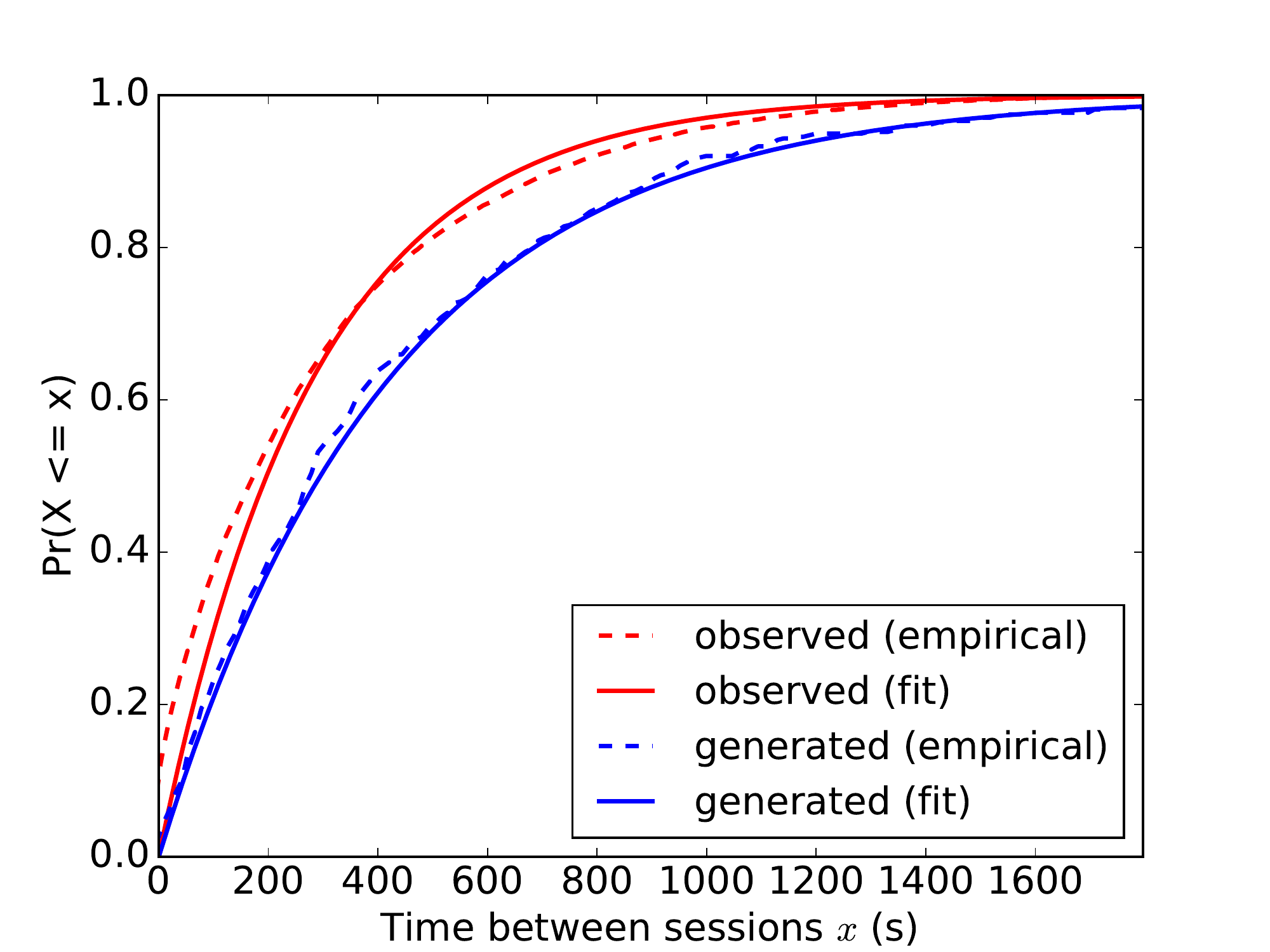}
	\label{fig:pavia_compare_tbs}} 
	\subfloat[WSU] {
		\includegraphics[width=0.25\textwidth]{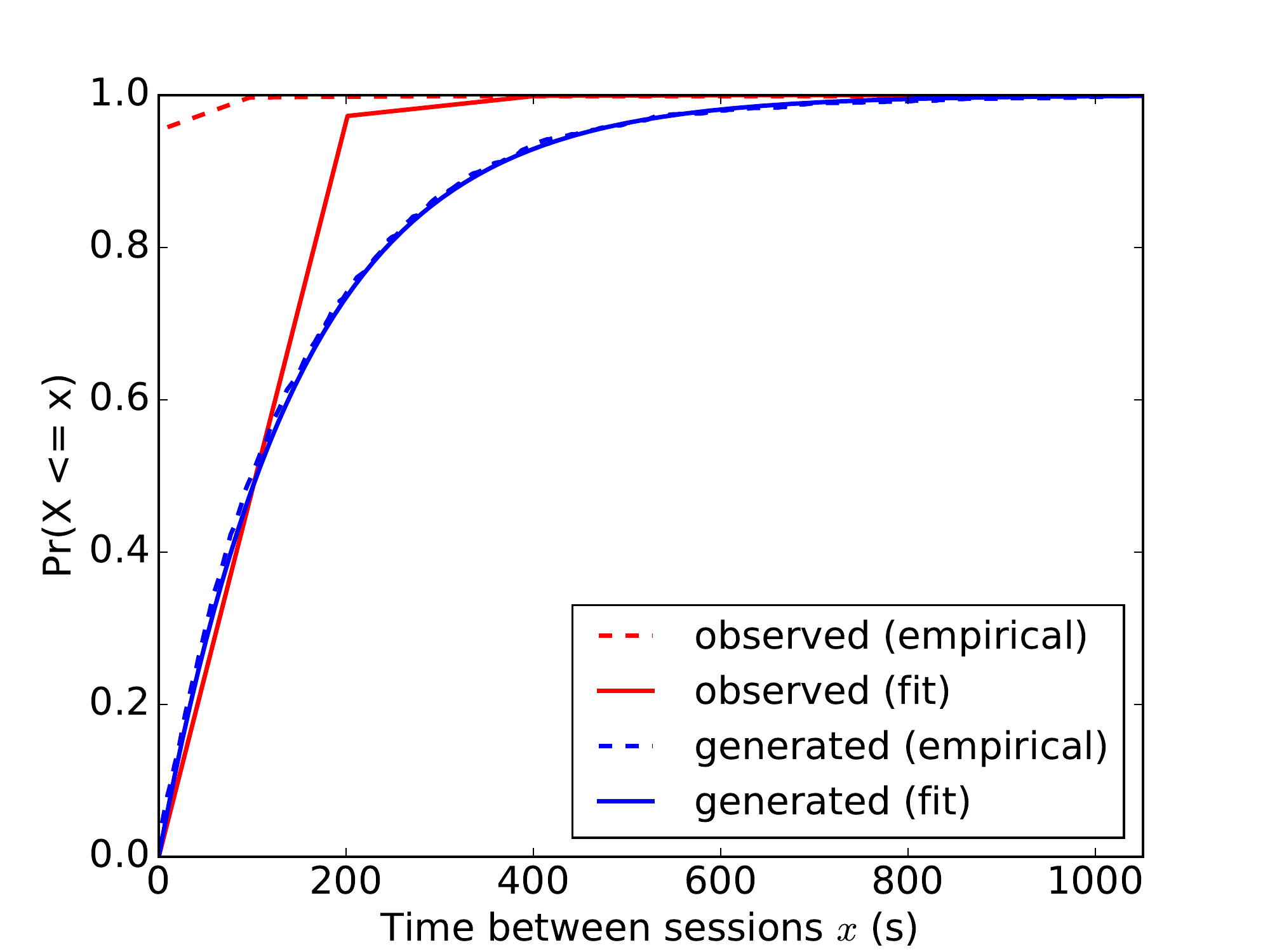}
	\label{fig:wsu_compare_tbs}} 
	\caption{CDFs of times between sessions}
	\label{fig:compare_tbs}

\end{figure}

Comparison of the time between the arrival of new sessions in
generated and actual data are shown in Figure~\ref{fig:compare_tbs}. Notice that in the case of
the Univ. of Pavia dataset, the original dataset approximately follows an 
exponential distribution. 
This observation further justifies the use of a Poisson process 
to generate the time between sessions as described in Section~\ref{sec:session_generation}.
The fit exponential is different between the actual and generated traffic. This is
because only a finite number of requests were generated, which can lead the fit
parameters of the generated data to differ from those of the original data. The interaction
of session arrival time generation with other components of the traffic generator could
also impact the resulting distribution. However, as can be seen from the figure,
the resulting empirical distribution is still quite close to the underlying exponential
distribution.

\subsection{Inter-Arrival Times within a Session}
\begin{figure}[t]
	\centering
	\subfloat[Univ. of Pavia dataset] {
		\includegraphics[width=0.25\textwidth]{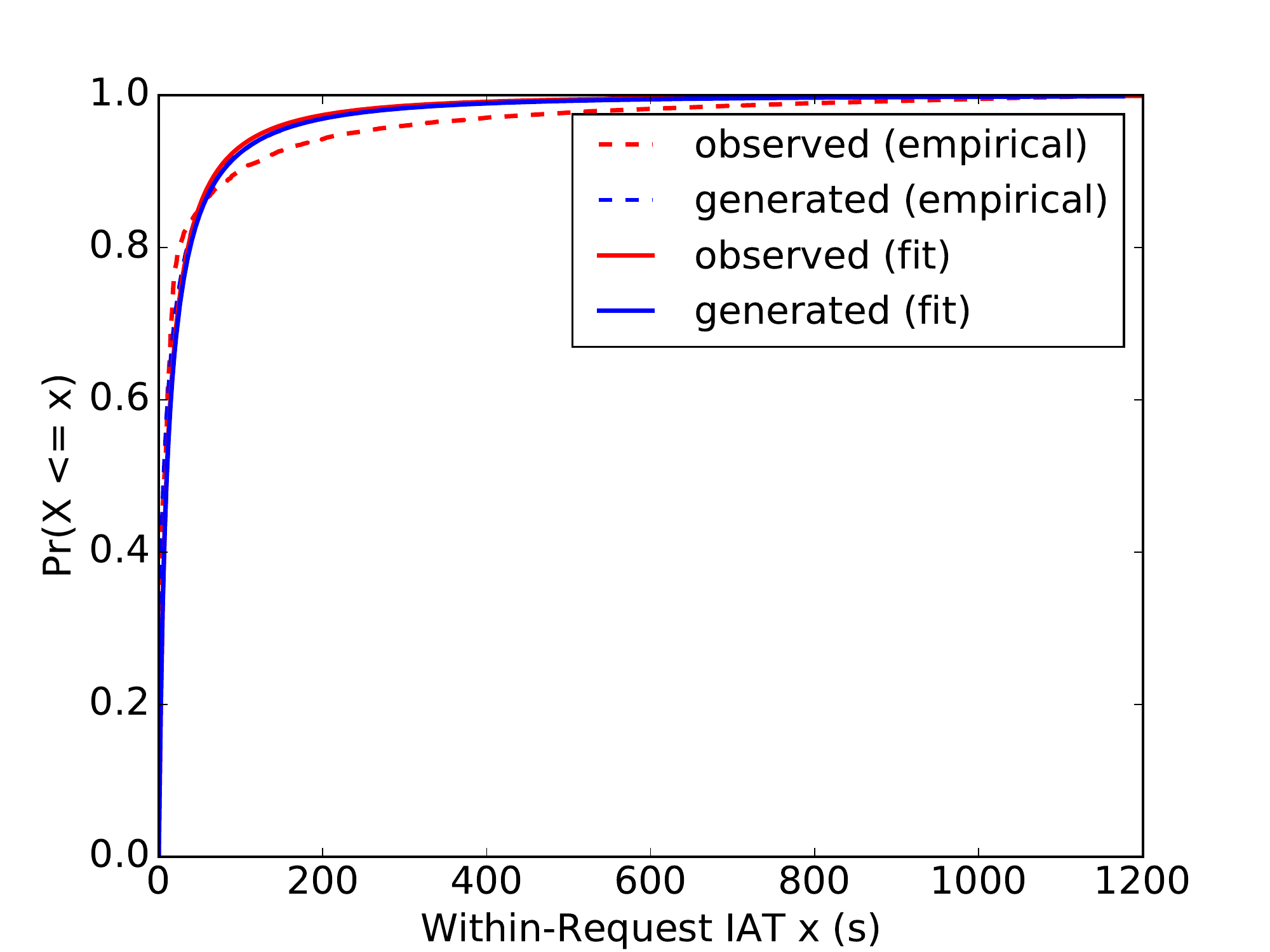}
	\label{fig:pavia_compare_iat}} 
	\subfloat[WSU dataset] {
		\includegraphics[width=0.25\textwidth]{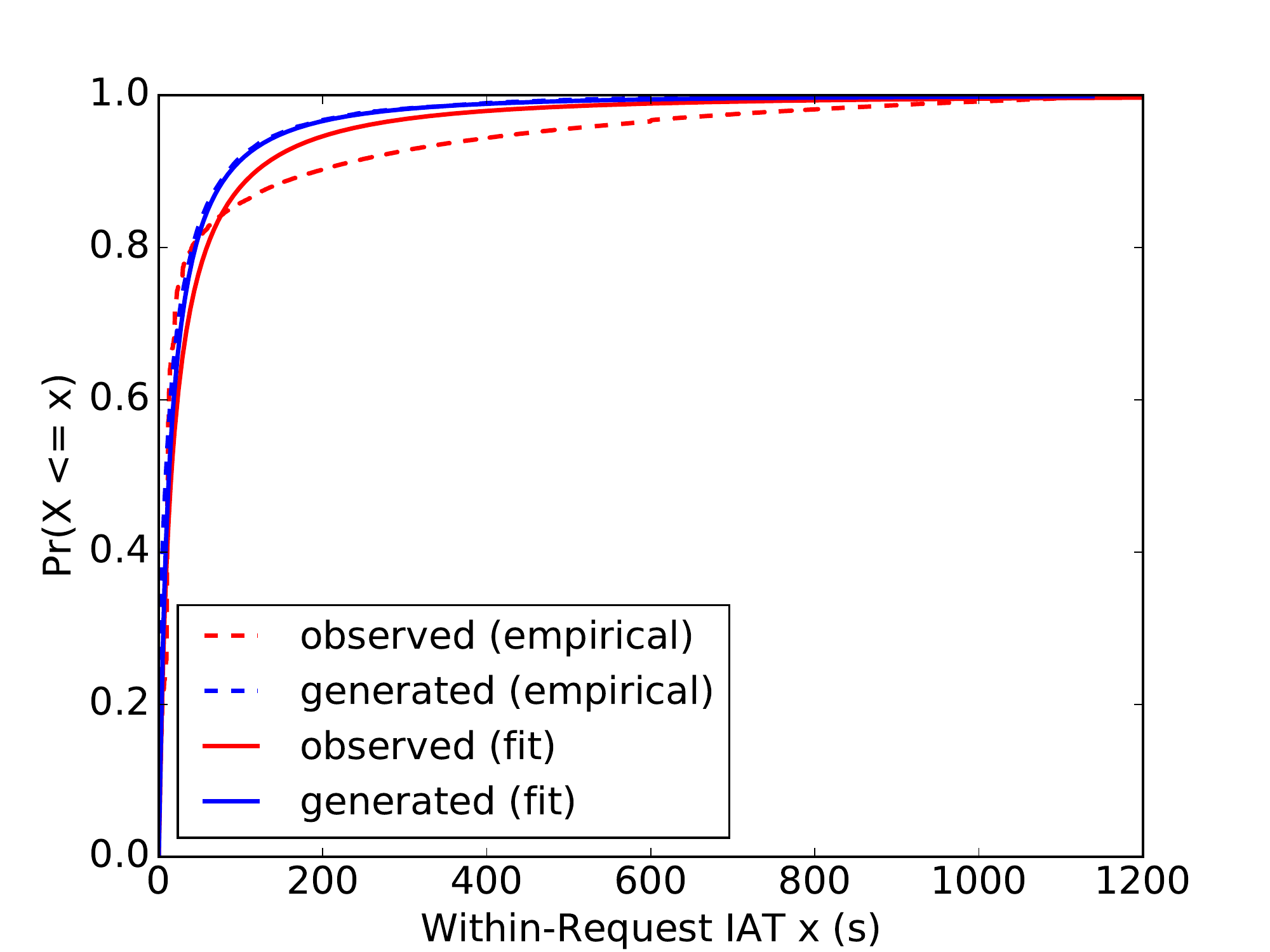}
	\label{fig:wsu_compare_iat}} 
	\caption{CDFs of inter-arrival times within a session}
	\label{fig:compare_iat}
	\vspace{-10px}
\end{figure}

Plots showing the empirical and theoretical cumulative distributions of the times
between requests within sessions are given in Figure~\ref{fig:compare_iat}. Recall
that a lognormal distribution was used to model these times. For the Univ. of
Pavia dataset, the fit distributions for generated and real traffic match quite closely.
The empirical distribution for the real traffic is not a close match to the fit, while
the generated traffic is. This is because the time between requests was generated using
a lognormal distribution, while in the original dataset a more complicated process,
something other than lognormal, could have led to the resulting empirical distribution.
While the differences between the fitted log-normals are more marked in the plot for the 
WSU dataset, these differences are not significant. In all cases, the
empirical distribution of the generated data matches its fit, indicating that other
components of the traffic generator have less impact on the time between requests
than they do, say, the time between sessions.

\subsection{Session Length}
\begin{figure}[t]
	\centering
	\vspace{-10px}
	\subfloat[Univ. of Pavia dataset] {
		\includegraphics[width=0.25\textwidth]{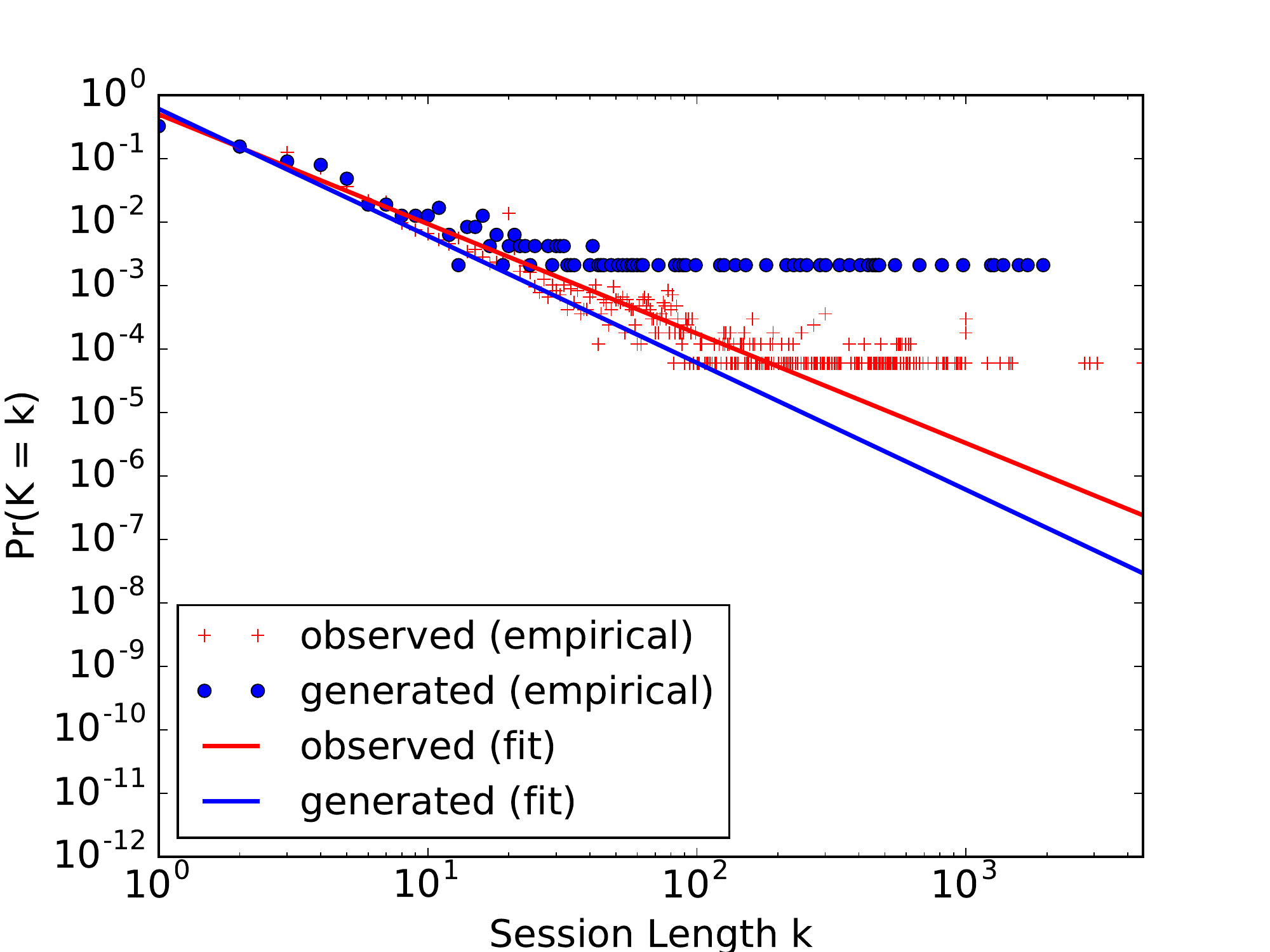}
	\label{fig:pavia_compare_sess_len}} 
	\subfloat[WSU dataset] {
		\includegraphics[width=0.25\textwidth]{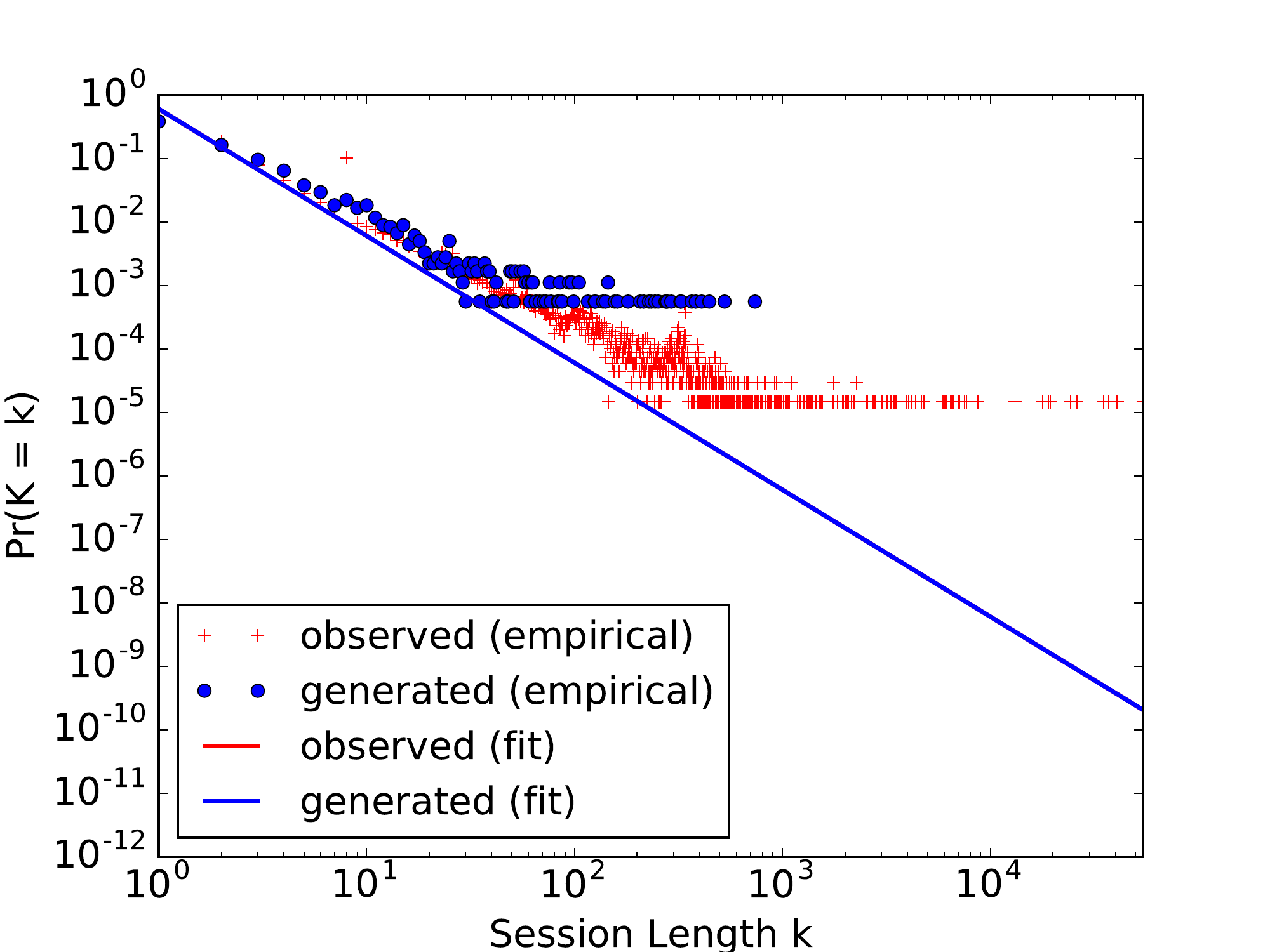}
	\label{fig:wsu_compare_sess_len}}
	\caption{PDFs of session length distributions}
	\label{fig:compare_sess_len}
\end{figure}

Log-log plots of the empirical and theoretical probability mass functions of session
lengths for generated and actual data are shown in Figure~\ref{fig:compare_sess_len}.
In the Univ. of Pavia dataset, the theoretical distributions differ in the tail,
possibly due to high variation in the tails making it difficult to achieve a good fit
to a power law. This variance can be seen in both generated and original data points,
and is a consequence of not having an extremely large number of observations so that
enough high values, which have a low but not negligible probability of occurring in
power-tailed distributions, are observed to make the power-tail behavior apparent
throughout the entire support.

\subsection{Cache Performance}
There are many ways that Web robot traffic could impact the performance of a Web server.
For example, a high volume of traffic could cause delays in server responses due to large
amount of threads being spun off on the server machine to process each request. A common
way to speed up server response is by caching commonly requested static content. 
Large
variations in the resources requested by Web robots can negatively impact cache performance.
In order to test new caching algorithms that may be able to handle Web robot traffic well,
large amounts of traffic, either observed or generated, are needed. By seeing if our
generated Web robot traffic impacts a hypothetical Web cache in the same way as actual
traffic, we can know if our generation system is useful in testing new caching algorithms.

\begin{figure}[t]
	\centering
	\subfloat[Univ. of Pavia dataset] {
		\includegraphics[width=0.25\textwidth]{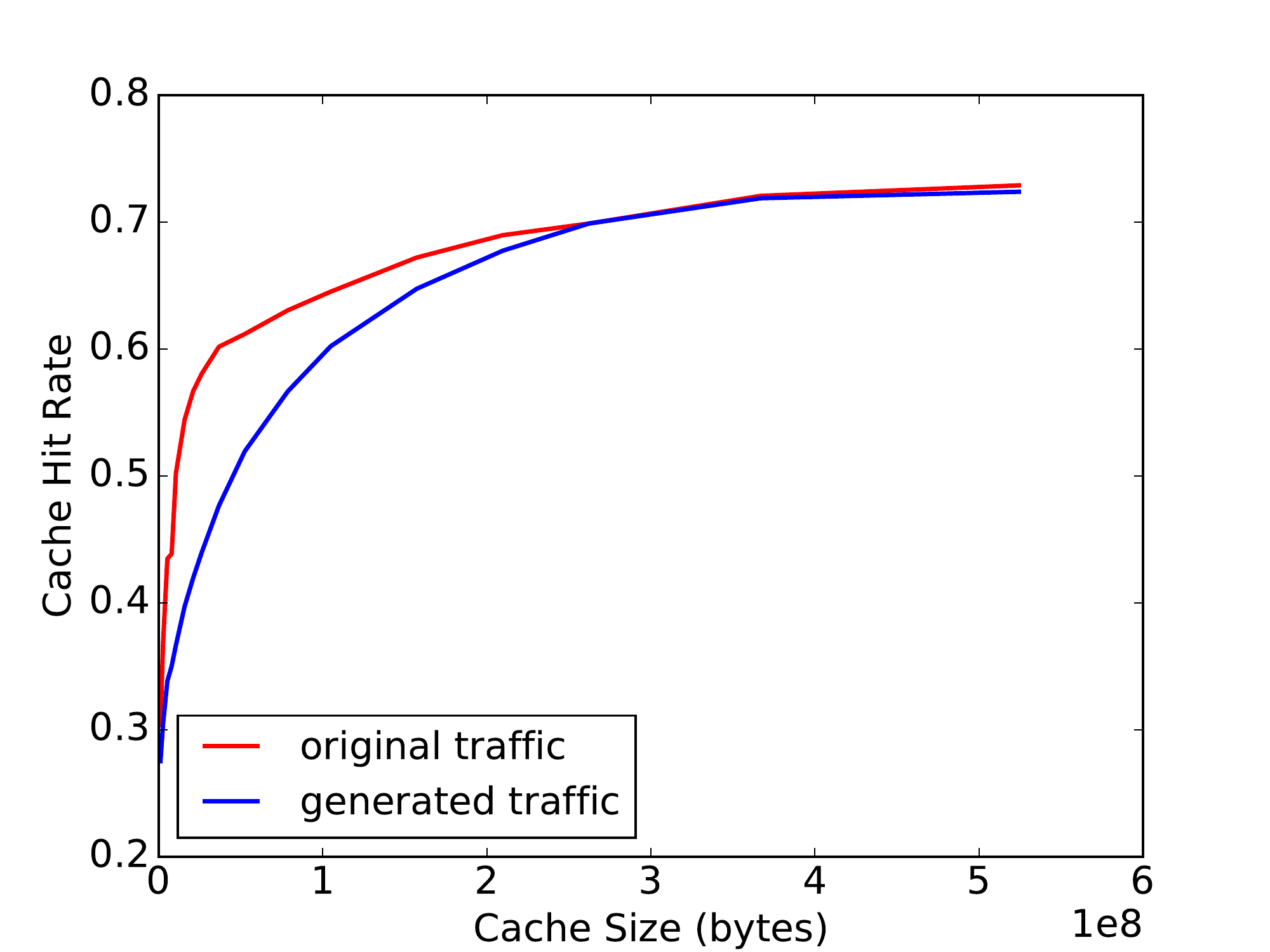}
	\label{fig:pavia_lfu_hit}}
	\subfloat[WSU dataset] {
		\includegraphics[width=0.25\textwidth]{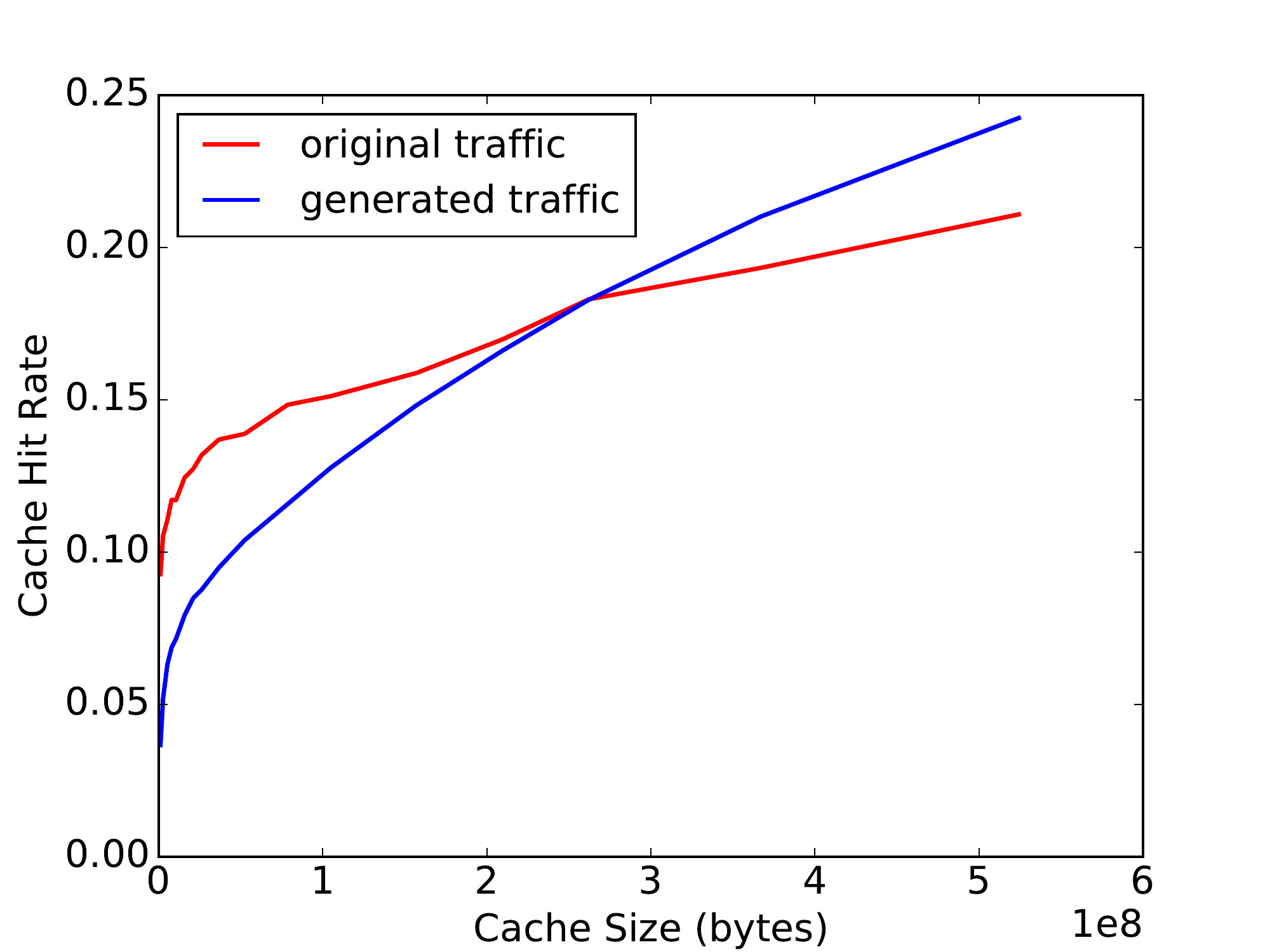}
	\label{fig:wsu_lfu_hit}}
	\caption{Simulated cache hit rates with LFU eviction policy}
	\label{fig:lfu_hit}
\end{figure}

\begin{figure}[t]
	\centering
		\vspace{-24px}
	\subfloat[Univ. of Pavia dataset] {
		\includegraphics[width=0.25\textwidth]{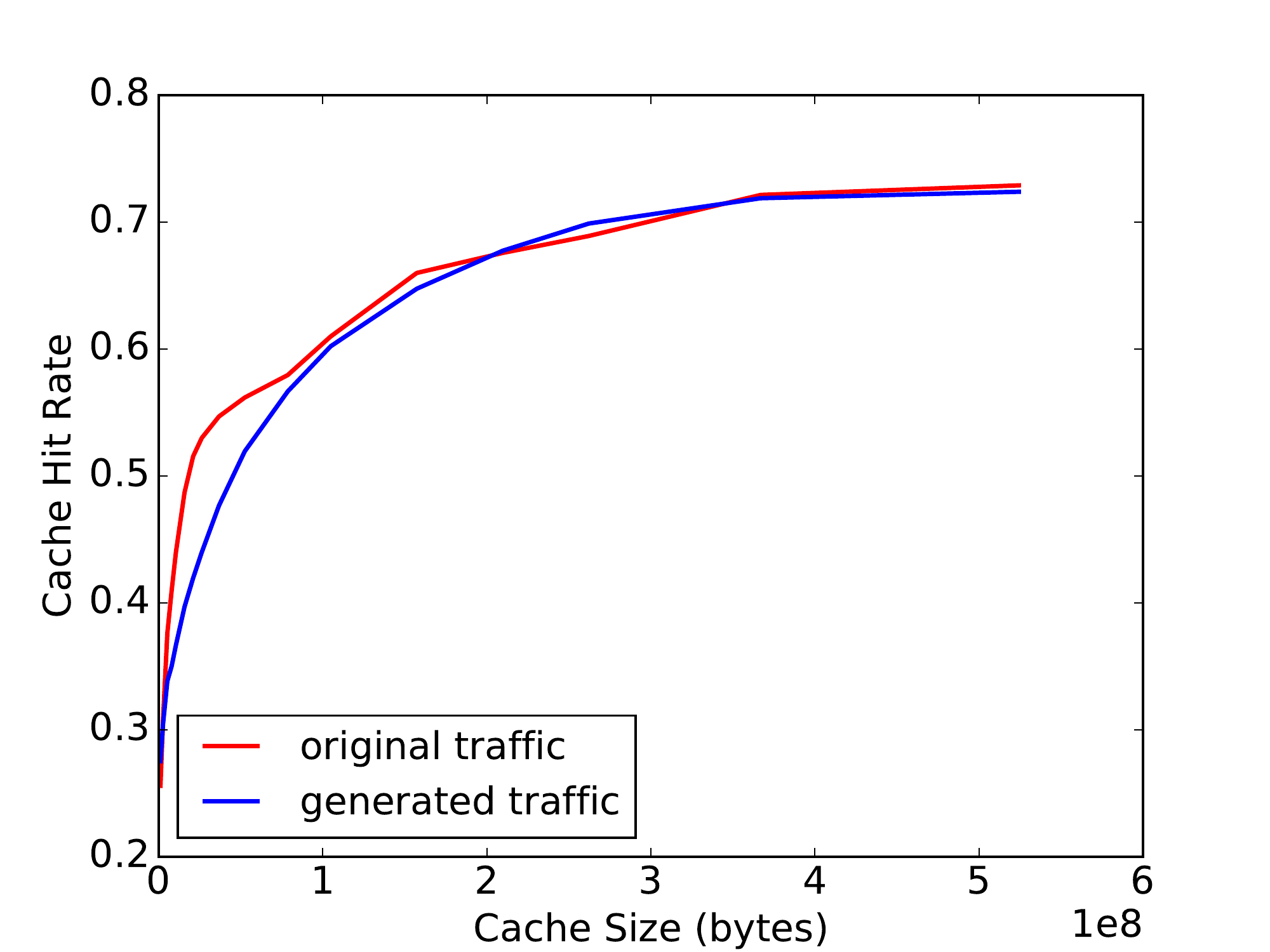}
	\label{fig:pavia_lru_hit}} 
	\subfloat[WSU dataset] {
		\includegraphics[width=0.25\textwidth]{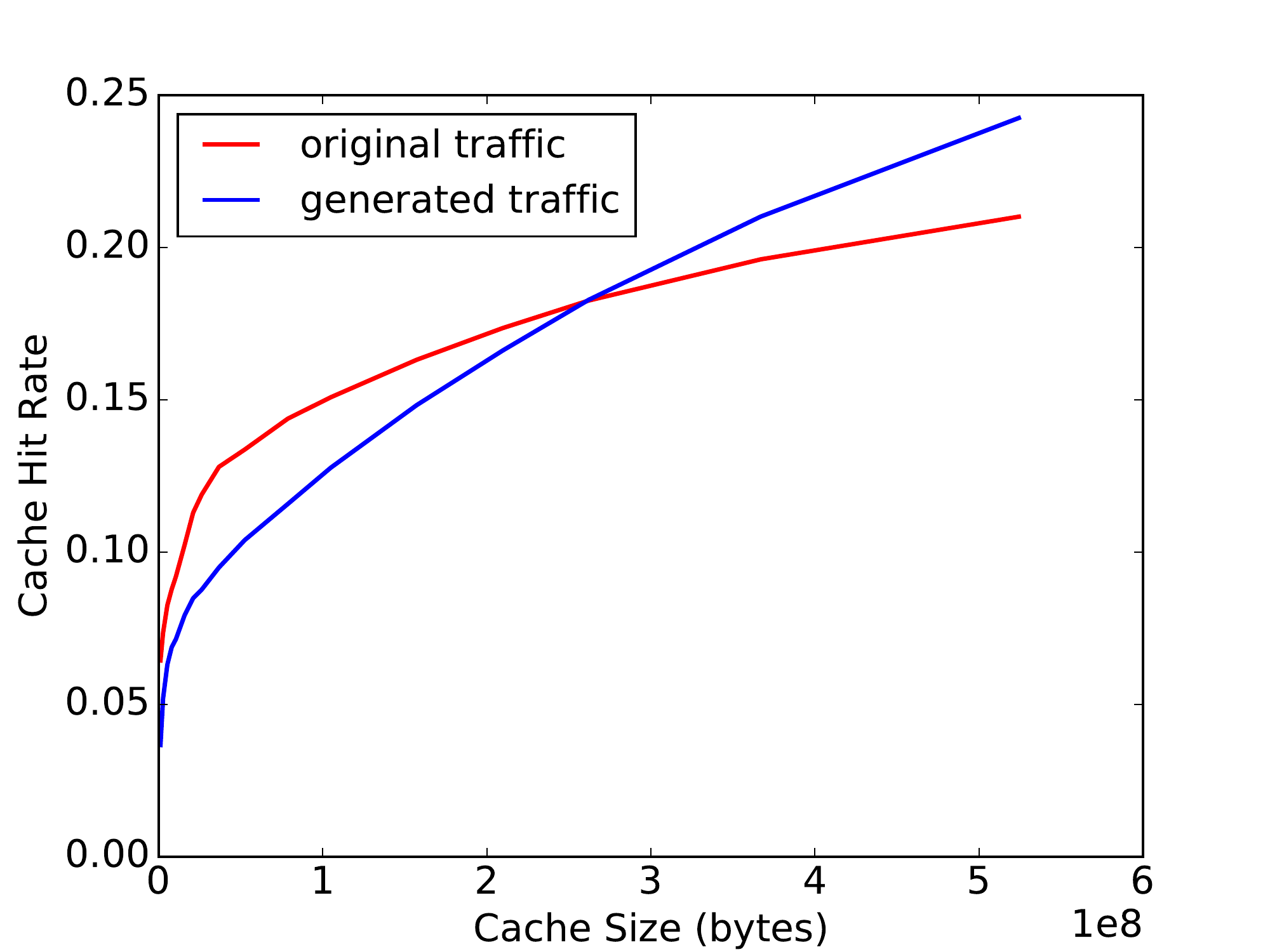}
	\label{fig:wsu_lru_hit}}
	\caption{Simulated cache hit rates with LRU eviction policy}
	\label{fig:lru_hit}

\end{figure}

To evaluate how our traffic generation affects Web server caches, we implemented a
simulation of a simple cache with two eviction policies: Least-frequently used (LFU),
which removes the object with the least amount of hits when the cache is full, and
least-recently used (LRU), which removes the object which has been in the cache the
longest. These and other cache replacement strategies are described in more detail 
in~\cite{podlipnig2003survey,wong2006web}. We then computed the average cache hit 
rate for varying cache sizes on generated and actual traffic, and plotted the results. 
The results are in Figure~\ref{fig:lfu_hit} for the LFU policy and 
Figure~\ref{fig:lru_hit} for the LRU policy.

The curves for the University of Pavia dataset closely match, meaning
that the traffic generator is able to reproduce characteristics of Web
robot traffic at that server which impact cache performance in the same
way as actual traffic. The generated LRU hit-rate curve is closer to
the actual curve than the LFU curve. This may be due to the fact that
the LRU policy is based on the time a resource was last requested, and
the traffic generator could be better at simulating the temporal aspects
of traffic. However, the curves for the WSU dataset
do not match as well. This indicates that there may be some unusual Web
robot behavior in the traffic which is not accounted for by the traffic
generator. In particular, there is some divergence in cache performance
for large cache sizes, compared to the University of Pavia curves which
grow closer as the cache size increases. Since the original data has
worse cache performance for large cache sizes, we hypothesize that this
is due to dynamic and new resources on the \url{wright.edu} domain, which
appear later in the original traffic but which may appear throughout the
entire generated traffic.

The hit-rate curves for generated data appear to be smoother than the
curves for actual data. This may be due to the fact that the generated data
maintains the same characteristics (session length etc.) throughout the
entire trace, while actual data may exhibit local differences in characteristics,
such as higher activity during the daytime, which can cause variations in
the hit-rate of a cache at differing times. The integration and modeling of these
local variations are left for future work.

\section{Conclusion}
\label{sec:conclusion}
In this paper we introduced a new system for generating Web robot requests 
which share some of the characteristics of actual Web robot traffic. We 
described how the temporal aspects of the traffic, such as the times between
session arrivals, are modeled by various probability distributions which can
be sampled from to generate a sequence of requests whose temporal characteristics
are similar to that of actual traffic. We then introduced methods for choosing
robots to assign sessions to, and generating a sequence of subdirectories to draw
resources from. For the actual generation of resources, we introduced a Bayesian
model, fit using MAP, which incorporates prior information from all robots in a
given subdirectory along with data from a single robot, to produce a model which
is influenced by all robots but is still unique for each robot. 

To evaluate our traffic simulator, we generated many requests and compared the
statistical properties of the results with those of the original traffic we 
were attempting to model. We used two datasets from different academic Web servers,
one located in the United States, and the other in Italy. The characteristics 
we used for comparison were time between sessions, inter-arrival times within 
a session, session length, and impact on a hypothetical Web server cache with 
LFU and LRU eviction policies. We showed that our generated traffic is able
to match the trends observed in the original datasets.

There are two main ways our traffic generation approach could be improved: we
could use better models, particularly for selecting the next active robot, and
choosing subdirectories and resources, or we could add completely new components
to our generator to capture characteristics of traffic which were previously
unnoticed. For selecting the next active robot, it might be better to take into
account any possible seasonality robots exhibit, i.e. the probability of a given
robot becoming active depends on the internal clock of the traffic generator. We
could choose the subdirectory and resource at the same time, based on which robot
is making the resource and the past requests of that robot. New components could
use measurements such as the size of resources, hyperlink structure, and types
of robots (Web crawler, link checker, etc.) to further improve our models and
the quality of the generated data.

Future studies could also look at datasets from commercial web servers, or any
other non-academic site. The traffic at these websites may have properties which
are not as adequately modeled by our system, and understanding them could lead to
future improvements in the generation of Web robot traffic. These studies could
also look at properties of the traffic we didn't consider; among these are the sizes
of resources, sequential structures (what resources are likely to
be requested in sequence?), and changes in all previous properties over time. 

\section*{Acknowledgments}
\label{sec:acknowledgment}
The authors thank Logan Rickert for data processing support,
Maria-Carla Calzarossa for data from the University of Pavia, and Mark 
Anderson for data from Wright State University. 
This paper is based on work supported by the National Science Foundation 
(NSF) under Grant No. 1464104. Any opinions, findings, and conclusions or 
recommendations 
expressed in this material are those of the author(s) and do not necessarily 
reflect the views of the NSF.

\bibliographystyle{IEEEtran}
\bibliography{cdmake17traffic}

\end{document}